\documentclass[letterpaper,10pt]{article}

\usepackage{amsmath}
\usepackage{amssymb}
\usepackage{psfrag}

\newcommand{\dd}{\textrm{d}}

\author{A. L\'opez-Ortega\thanks{alopezo@ipn.mx} \\
Centro de Investigaci\'on en Ciencia Aplicada y Tecnolog\'{\i}a Avanzada. \\ 
Unidad Legaria. Instituto Polit\'ecnico Nacional. \\
Calzada Legaria \# 694. Colonia Irrigaci\'on. Delegaci\'on Miguel Hidalgo. \\
M\'exico, D.\ F., M\'exico. \\
C.\ P.\  11500  
}

\title{Entropy spectra of single horizon black holes in two dimensions}

\begin{document}

\maketitle

\begin{abstract}

The Hod conjecture proposes that the asymptotic quasinormal frequencies determine the entropy quantum of a black hole. Considering the Maggiore modification of this conjecture we calculate the entropy spectra of general, single horizon, asymptotically flat black holes in two-dimensional dilaton gravity. We also compute the entropy quanta of the two-dimensional Witten and AdS$_2$ black holes. Using the results for the entropy quanta of these two-dimensional black holes we  discuss whether the produced values are generic. Finally we extend the results on the entropy spectra of other black holes.

\vspace{.3cm}


\end{abstract}

\section{Introduction}
\label{s: Introduction}

For a stationary black hole, Bekenstein \cite{Bekenstein:1974jk}--\cite{Bekenstein:1998aw} suggests that in the semiclassical limit the area spectrum of its event horizon takes the form\footnote{In units where $G=k_B=c=1$. Here $G$ is the gravitational constant, $k_B$ is the Boltzmann constant, and $c$ is the velocity of light in vacuum.}
\begin{equation} \label{e: area spectrum Bekenstein}
 A_s \approx \epsilon \hbar s , \qquad \qquad s=0,1,2,\dots,
\end{equation} 
with $\hbar$ denoting the reduced Planck constant and $\epsilon$ is a parameter of order unity. We believe that the exact value of $\epsilon$ must be determined by the quantum theory of gravity, nevertheless, based on some assumptions, the parameter $\epsilon$ is calculated in several references  \cite{Bekenstein:1995ju}--\cite{Maggiore:2007nq}, and we notice that two values of $\epsilon$ appear frequently, $\epsilon = 4 \ln (3)$ and $\epsilon = 8 \pi$. 

One way to get the value $\epsilon = 4 \ln (3)$ is to assume a strict statistical interpretation of the black hole entropy, in such a way that the area elements of the event horizon have three internal degrees of freedom \cite{Bekenstein:1995ju,Mukhanov:1986me}. For the four-dimensional Schwarzschild black hole the results obtained with Hod's conjecture support this value of $\epsilon$ (see Ref.\ \cite{Hod:1998vk}). According to this proposal, the real part of the asymptotic quasinormal frequencies (AQNF in what follows) determines the change of energy in a transition between two quantum levels of the black hole \cite{Hod:1998vk}. The proposed relation is 
\begin{equation} \label{e: Hod conjecture formula}
 \Delta E = \lim_{n \to \infty} \hbar \omega_{R,n},
\end{equation}  
where $\Delta E$ is the change in the energy of the black hole, $\omega_{R,n}$, denotes the real parts of the AQNF, and $n$ is the mode number. 

For the four-dimensional Schwarzschild black hole of mass $M$ the AQNF of its gravitational perturbations are \cite{Hod:1998vk},  \cite{Nollert:1993zz}--\cite{Natario:2004jd} 
\begin{equation} \label{e: Schwarzschild AQNF 4D}
 \omega = \frac{\ln (3)}{8 \pi M } - \frac{i}{4 M}\left( n + \frac{1}{2} \right), \qquad n \in \mathbb{N}, \quad n \to \infty .
\end{equation} 
From the formula (\ref{e: Hod conjecture formula}), Hod finds that the area spectrum of the four-dimensional Schwarzschild black hole is of the form (\ref{e: area spectrum Bekenstein}) with $\epsilon = 4 \ln (3)$ \cite{Hod:1998vk}.

Although for the four-dimensional Schwarzschild black hole Hod's conjecture produces a reasonable value of $\epsilon$ when we use the AQNF of the gravitational perturbations (\ref{e: Schwarzschild AQNF 4D}), the conjecture encounters difficulties when it is used to calculate the area spacing of the event horizons for other spacetimes \cite{Maggiore:2007nq,Setare:2004uu,LopezOrtega:2009ww}. For example, for some black holes we expect a discrete and evenly spaced area spectrum \cite{Bekenstein:1974jk,Bekenstein:1998aw,Gour:2002ga,Gour:2002uk}, nevertheless Hod's conjecture yields a discrete area spectrum but it is not evenly spaced \cite{Setare:2004uu}. Another example is the $D$-dimensional de Sitter spacetime, since for the gravitational perturbations their quasinormal frequencies (QNF in what follows) have real parts equal to zero \cite{Natario:2004jd,LopezOrtega:2006my}, and the result that yields Hod's conjecture for its area quantum is not easy to understand \cite{LopezOrtega:2009ww}. Moreover, for the four-dimensional Schwarzschild black hole, Hod's conjecture predicts that the value of the area quantum depends on the field exploited in its calculation \cite{Maggiore:2007nq}.

Following Ref.\ \cite{Kettner:2004aw}, we believe that a requirement to  seriously consider Hod's conjecture is to study whether it gives general results for the entropy spectra; nevertheless the real parts of the AQNF of distinct black holes may be different; furthermore for some black holes the real parts of their AQNF may depend on the field. Hence the original Hod conjecture produces different values of the entropy quanta \cite{Hod:1998vk,Maggiore:2007nq}. Motivated by these facts, for a number of spacetimes as large as possible, it is convenient to study the results for the entropy quantum that yields the Hod conjecture and the modification proposed by Maggiore \cite{Maggiore:2007nq} (the Hod-Maggiore proposal in what follows). Hence using the Hod and the Hod-Maggiore proposals we calculate the entropy quanta of several two-dimensional black holes and discuss the differences between the values given by the two proposals.

The present paper is organized as follows. In Section \ref{s: generic 2D black holes}, using the Hod-Maggiore proposal, we calculate the entropy quanta of a family of two-dimensional black holes for which we know the AQNF of the Klein-Gordon field. We also compute the entropy spectra of the two-dimensional Witten and AdS$_2$ black holes based on the exact QNF of the Klein-Gordon and Dirac fields. In Section \ref{s: Some extensions} we discuss some relevant facts on the original Hod conjecture and the Hod-Maggiore proposal. Furthermore in this section we extend the results on the entropy quanta of other spacetimes as Schwarzschild, de Sitter, and BTZ. Finally in Section \ref{s: Summary} we present a brief summary of our results.

\section{Single horizon two-dimensional black holes}
\label{s: generic 2D black holes}

As it is well known, two-dimensional gravity allows us to study in a simple setting many physical phenomena \cite{Grumiller:2002nm}. For example, based on Hod's conjecture in Ref.\ \cite{Kettner:2004aw} it is calculated the entropy quanta for a family of single horizon two-di\-men\-sion\-al black holes in order to determine whether this conjecture gives one or many values for them. Since our aim is to compute and analyze the entropy quanta that predicts the Hod-Maggiore conjecture \cite{Maggiore:2007nq} for these two-dimensional black holes, here we expound the results of Ref.\ \cite{Kettner:2004aw} that we shall use in what follows.

Following Ref.\ \cite{Kettner:2004aw} we consider the two-dimensional action
\begin{equation} \label{e: action gravity two}
 I_{2D} = \frac{1}{2 G} \int \dd^2 x \sqrt{|g|} \left( \phi R + \frac{V_{\phi}(\phi)}{\lambda^2} \right) ,
\end{equation} 
with $R$ denoting the scalar curvature, $\phi$ the dilaton field, $\lambda$ a length scale, and in what follows $V_{\phi}$ is a power law potential of the form  $V_{\phi} = (1-b) \phi^{-b}$. It is known that a family of single horizon black holes is solution to the equations of motion derived from the action (\ref{e: action gravity two}) with a power law potential \cite{Kettner:2004aw,Cadoni:1995dd}. In the gauge $\phi = x / \lambda ,$  this family of black holes is given by \cite{Kettner:2004aw,Cadoni:1995dd} 
\begin{align} \label{e: black holes 2D}
\dd s^2  = -f(x) \dd t^2 + f^{-1}(x) \dd x^2, \qquad f(x)  = \left( \frac{x}{\lambda}\right)^{1-b} - 2 \lambda G M,
\end{align} 
where the parameter $b$ satisfies $-1 \leq b < 1$.

In a similar way to Ref.\ \cite{Kettner:2004aw}, we restrict the analysis to the interval $0 < b < 1 $, and in what follows when we refer to the single horizon black holes (\ref{e: black holes 2D}), implicitly we assume this restricted range for the parameter $b$. Notice that we impose the condition $b>0$  to get asymptotically flat black holes.

In a similar way to Ref.\ \cite{Kettner:2004aw} we consider a Klein-Gordon field $\Phi$ coupled to the dilaton that satisfies the equation of motion 
\begin{equation} \label{e: Klein-Gordon equation coupled}
 \partial_\mu(\sqrt{-g} h(\phi) g^{\mu \nu} \partial_\nu \Phi) = 0 ,
\end{equation} 
where, $h(\phi) = \phi^a$, $a$ is a continuous parameter, $a>0$, in such a way that the function $h(\phi)$ determines the coupling between the Klein-Gordon field and the dilaton. When we use the tortoise coordinate near the origin, the parameter $a$ sets the coupling between the perturbing Klein-Gordon field and the gravitational field \cite{Kettner:2004aw}. 

In the black holes (\ref{e: black holes 2D}) the AQNF of the Klein-Gordon field (\ref{e: Klein-Gordon equation coupled}) are equal to (see the formula (36) of Ref.\ \cite{Kettner:2004aw})
\begin{equation} \label{e: AQNF 2D black holes}
\omega = \frac{\kappa}{2 \pi} \ln[1 + 2 \cos(\pi(a-1))] - i \kappa \left( n + \frac{1}{2} \right), \qquad n \in \mathbb{N}, \quad n \to \infty,
\end{equation} 
where 
\begin{equation} \label{e: surface gravity 2D}
 \kappa = \frac{|f^\prime (x_h)|}{2} =\frac{1}{2 \lambda} V_{\phi} (\phi_h),
\end{equation} 
is the surface gravity of the black holes (\ref{e: black holes 2D}), $\phi_h$ denotes the value of the dilaton field at the event horizon, and $x_h$ denotes the radius of the event horizon. Note that the AQNF (\ref{e: AQNF 2D black holes}) do not depend on the parameter $b$, but they are valid only for the black holes whose parameter $b$ satisfies $0 < b < 1$.

To get the AQNF (\ref{e: AQNF 2D black holes}), in Ref.\ \cite{Kettner:2004aw} the behavior of the coupled Klein-Gordon field near the origin is used in the calculation, therefore their real parts depend on the parameter $a$. We point out that due to the behavior  near the event horizon of the tortoise coordinate for the black holes (\ref{e: black holes 2D}), the surface gravity appears in the imaginary parts of the AQNF (\ref{e: AQNF 2D black holes}) because in the method of Ref.\ \cite{Kettner:2004aw} a contour integral around the event horizon is computed.

From the AQNF (\ref{e: AQNF 2D black holes}), it follows that in their real parts the argument of the logarithm is an integer only in exceptional cases, in disagreement with the integral values that are necessary for a strict statistical interpretation of the black hole entropy \cite{Bekenstein:1995ju,Mukhanov:1986me}. Therefore for the black holes (\ref{e: black holes 2D}) Hod's conjecture yields that their entropy quanta depend on the parameter $a$, and in this framework it is not an easy task to find a physical interpretation to the dependence on the parameter $a$ of their entropy quanta \cite{Kettner:2004aw}.

Considering that the quasinormal modes (QNM in what follows) of a black hole behave as the oscillations of a damped oscillator, based on Hod's conjecture, and in order to overcome some of its difficulties, in Ref.\ \cite{Maggiore:2007nq} Maggiore proposes that the entropy spectrum is determined by the asymptotic value of the physical frequencies defined by \cite{Maggiore:2007nq}
\begin{equation} \label{e: definition physical frequency} 
 \omega_{p,n}= \lim_{n \to \infty} \sqrt{\omega_{R,n}^2 + \omega_{I,n}^2},
\end{equation} 
with $\omega_{I,n}$ denoting the imaginary parts of the QNF. Recently this proposal has attracted attention and its results for the entropy spectra of several black holes are presented in Refs.\ \cite{Maggiore:2007nq}, \cite{LopezOrtega:2009ww}, \cite{Wei:2009yj}--\cite{LopezOrtega:2010tg}.

As far as we know in the framework of the Hod-Maggiore proposal the entropy quanta for the family of black holes (\ref{e: black holes 2D}) are not calculated in detail. We also consider other two black holes in two dimensions whose parameter $b$ is outside the interval $0<b<1$. We point out that based on the Hod-Maggiore proposal there are two methods to find the entropy quanta of the black holes \cite{Kunstatter:2002pj,Maggiore:2007nq,Vagenas:2008yi}. Here we use both methods to calculate the entropy quanta of the black holes (\ref{e: black holes 2D}) and discuss the obtained results. In what follows we often identify the change in the horizon entropy due to the emission of a quantum with the entropy quantum of the horizon \cite{Hod:1998vk,Maggiore:2007nq}. 

\textbf{ \textit{First method:}} This method is based on the ideas by Hod \cite{Hod:1998vk} and Maggiore \cite{Maggiore:2007nq}. According to Maggiore the emission of a quantum yields that the energy of the black hole changes in discrete steps determined by (compare with the formula (\ref{e: Hod conjecture formula}))
\begin{equation} \label{e: delta E}
 \Delta E = \hbar \, \Delta \omega_p,
\end{equation} 
where
\begin{equation} \label{e: physical frequency}
 \Delta \omega_p = \lim_{n \to \infty} (\omega_{p,n+1} - \omega_{p,n}). 
\end{equation} 
From the AQNF (\ref{e: AQNF 2D black holes}) we get that the quantity $\Delta \omega_p$ is equal to
\begin{equation} \label{e: delta omega 2D}
 \Delta \omega_p = \kappa.
\end{equation} 

For the black holes (\ref{e: black holes 2D}) their entropies $S$ are equal to \cite{Gegenberg:1994pv}
\begin{equation}  \label{e: entropy 2D black hole}
 S = \frac{2 \pi}{\hbar G} \phi_h ,
\end{equation} 
and therefore we get 
\begin{equation} \label{e: delta S}
 \Delta S = \frac{2 \pi }{G \hbar} \Delta \phi .
\end{equation} 
Taking into account that for the two-dimensional black holes the variation of their ADM energies under small perturbations is determined by (see the formula (39) of Ref.\ \cite{Gegenberg:1994pv}\footnote{Our potential $V_{\phi}$ is the negative of the potential used by Gegenberg et al.\ in Ref.\  \cite{Gegenberg:1994pv}.})
\begin{equation} \label{e: delta E 2D black holes}
 \Delta E  = \frac{1}{2 \lambda G} V_{\phi} (\phi_h) \Delta \phi ,
\end{equation} 
and from the formulas (\ref{e: surface gravity 2D}), (\ref{e: delta E}), (\ref{e: delta omega 2D}), (\ref{e: delta S}) we get that for the entropy quanta of the black holes (\ref{e: black holes 2D}) the Hod-Maggiore proposal yields a unique value given by
\begin{equation} \label{e: entropy quantum 2D}
 \Delta S = 2 \pi.
\end{equation}  

For the black holes (\ref{e: black holes 2D}) we notice that the entropy quantum (\ref{e: entropy quantum 2D}) is independent of the parameter $a$, in contrast to original Hod's conjecture that predicts entropy quanta that depend on this parameter \cite{Kettner:2004aw}. Furthermore the entropy quantum (\ref{e: entropy quantum 2D}) of the black holes (\ref{e: black holes 2D}) is equal to that of the $D$-dimensional Schwarzschild black hole ($D \geq 4$)  \cite{Maggiore:2007nq,Wei:2009yj,LopezOrtega:2010tg},  four-di\-men\-sion\-al Kerr black hole \cite{Vagenas:2008yi,Medved:2008iq,Kwon:2011da,Myung:2010af,Kwon:2010mt}, and $D$-dimensional Reissner-Nordstr\"om black hole \cite{Kwon:2011da,Kwon:2010mt,LopezOrtega:2010tg}. Furthermore Medved proposes that this value for the entropy quantum is universal \cite{Medved:2009nj}--\cite{Medved:2004mh}.

\textbf{ \textit{Second method:} }  This method is based on the ideas by Kunstatter \cite{Kunstatter:2002pj} and Maggiore \cite{Maggiore:2007nq}. The method is proposed and used  with the original Hod conjecture in Ref.\ \cite{Kunstatter:2002pj}; here we exploit it, with the modifications suggested in Refs.\  \cite{Maggiore:2007nq}, \cite{Vagenas:2008yi}, \cite{Medved:2008iq}, to calculate the entropy spectra of the black holes (\ref{e: black holes 2D}). 

Kunstatter \cite{Kunstatter:2002pj} notices that for a system of energy $E$ and oscillation frequency $\omega$, the integral
\begin{equation} \label{e: adiabatic invariant}
 I = \int \frac{\dd E}{\omega},
\end{equation} 
is an adiabatic invariant. According to the Bohr-Sommerfeld quantization, in the semiclassical limit the adiabatic invariant $I$ has an evenly spaced spectrum, so in this limit $I = n \hbar$ \cite{Kunstatter:2002pj}.

As we commented, Maggiore suggests that for the physical phenomena that we are interested in, the change in the parameters of the black hole are due to the transition of the type $\omega_{p,n+1} \to \omega_{p,n}$, and therefore the appropriate oscillation frequency of the two-dimensional black holes is the quantity $\Delta \omega_p$. Thus for these black holes the adiabatic invariant (\ref{e: adiabatic invariant}) transforms into  \cite{Vagenas:2008yi,Medved:2008iq}
\begin{equation} \label{e: adiabatic invariant 2D}
 I = \int \frac{\dd E}{\Delta \omega_p}.
\end{equation} 

Considering that the Hawking temperature of the black holes (\ref{e: black holes 2D}) is \cite{Gegenberg:1994pv} 
\begin{equation} \label{e: Hawking temperature}
 T=\frac{\hbar \kappa}{2 \pi} ,
\end{equation} 
from the formulas (\ref{e: delta omega 2D}), (\ref{e: Hawking temperature}), and the first law of the thermodynamics \cite{Grumiller:2002nm}, we find that the adiabatic invariant (\ref{e: adiabatic invariant 2D}) becomes
\begin{equation} \label{e: integral entropy}
 I = \int \frac{\dd E}{\kappa} = \frac{\hbar}{2 \pi} \int \dd S = \frac{ \hbar S}{2 \pi},
\end{equation} 
and hence, in the semiclassical limit, for the black holes (\ref{e: black holes 2D})  we get the entropy spectrum
\begin{equation} \label{e: spectrum discrete}
 S_n = 2 \pi n.
\end{equation} 

From the last formula we get that for the black holes (\ref{e: black holes 2D}) their entropy quanta  are given by the formula (\ref{e: entropy quantum 2D}). Thus both methods produce the same value for the entropy quanta of the black holes (\ref{e: black holes 2D}), in a similar way to several spacetimes that have been studied \cite{LopezOrtega:2009ww,Vagenas:2008yi,Medved:2008iq,LopezOrtega:2010tg}.

Notice that the first method produces the value of the entropy quantum, whereas the second method yields the mathematical form of the entropy spectrum and from this expression we derive the value of the entropy quantum.

It is convenient to note that the entropy spectrum (\ref{e: spectrum discrete}) of the black holes (\ref{e: black holes 2D}) is derived  with a different method in Ref.\ \cite{Barvinsky:1996hr}. To obtain this spectrum they quantized the Euclidean two-dimensional black holes (see also Ref.\ \cite{Barvinsky:2000gf}).
   
Although for the black holes (\ref{e: black holes 2D}) the Hod-Maggiore proposal produces a unique value for their entropy quanta, we note that this proposal does not give a universal value for the entropy quanta. For example, for the $D$-dimensional massless topological black hole \cite{LopezOrtega:2010wx} (MTBH in what follows) and the static three-dimensional BTZ black hole \cite{Kwon:2010um}, the Hod-Maggiore proposal yields an entropy quantum equal to $\Delta S = 4 \pi$ (which is identical to that for the $D$-dimensional de Sitter horizon \cite{LopezOrtega:2009ww}) whereas for Schwarzschild, Kerr, Reissner-Nordstr\"om, and black holes (\ref{e: black holes 2D}) it produces $\Delta S = 2 \pi$. 

For the black holes of general relativity with cosmological constant, the more puzzling result of the Hod-Maggiore proposal for an entropy quantum is that for the $D$-dimensional Schwarzschild anti-de Sitter black hole \cite{Daghigh:2008jz}. Daghigh and Green \cite{Daghigh:2008jz} show that for this black hole the entropy quantum predicted by the Hod-Maggiore proposal is $\Delta S = 4 \pi \sin( \pi/(D-1))$, and hence it depends on the spacetime dimension in a complicated way (see also Refs.\ \cite{Rong:2010}, \cite{Wei:2009sw}). It is convenient to note that for the enumerated spacetimes, their entropy quanta are independent of the cosmological constant.

Furthermore in Refs.\ \cite{Medved:2009nj}--\cite{Medved:2004mh} Medved suggests that $\Delta S = 2 \pi $ is the universal value for the entropy quantum, but as we have seen, even in general relativity, the Hod-Maggio\-re proposal produces values for the entropy quanta different from $\Delta S = 2 \pi $ in some spacetimes \cite{LopezOrtega:2009ww,Daghigh:2008jz,Kwon:2010um,LopezOrtega:2010wx}. We believe that this issue deserves further research.

To extend these results, in what follows we calculate the entropy spectra of the two-dimensional Witten \cite{Witten:1991yr,Mandal:1991tz} and AdS$_2$ black holes \cite{Achucarro:1993fd}. Our main reason is that for these two black holes the parameter $b$ is outside the interval $0<b<1$, because for a conformally related spacetime to the Witten black hole the parameter $b$ is equal to zero, whereas for the AdS$_2$ black hole we find that $b=-1$  \cite{Kettner:2004aw}. Furthermore for these two black holes their QNF are known in exact form \cite{Becar:2007hu}--\cite{ALO-RCE-IVA}.

Let us begin with the Witten black hole whose metric we take as \cite{Witten:1991yr,Mandal:1991tz}
\begin{align} \label{e: metric Witten}
 \dd s^2  = - \left(1-e^{-x}\right)\dd t^{2} + \frac{\dd x^{2}}{\left(1-e^{-x}\right)}.
\end{align} 
For this black hole the QNF of test Klein-Gordon and Dirac fields are calculated exactly in Refs.\ \cite{Becar:2007hu}--\cite{LopezOrtega:2011sc}. For the Klein-Gordon field of mass $m$ and coupled to scalar curvature with coupling constant $\zeta$ the QNF are \cite{Becar:2007hu,LopezOrtega:2009zx} 
\begin{equation} \label{e: QNF 2D Witten KG}
\omega = -\frac{i}{4} \left[ 2n + 1 \pm  \sqrt{1-4\zeta} - \frac{4 m^2}{2n + 1 \pm \sqrt{1 - 4 \zeta}}\right] , \qquad n=0,1,2,\dots,
\end{equation} 
when $\zeta \leq 1/4$ and
\begin{equation} \label{e: QNF 2D Witten KG 1}
\omega = \pm \sqrt{4 \zeta - 1} \frac{n^2 + n + \zeta + m^2}{4 n^2 + 4 n + 4 \zeta } - i \frac{(2 n + 1)(n^2 + n + \zeta - m^2)}{4 n^2 + 4 n + 4 \zeta},
\end{equation} 
when $\zeta > 1/4$. Moreover, the Dirac field with mass $m$ has QNF equal to \cite{LopezOrtega:2011sc}
\begin{align} \label{e: QNF Witten  Dirac 1}
\omega_{1} = -\frac{i}{2(n+1)} \left( (n+1)^{2} - m^{2} \right),  \qquad 
\omega_{1} = -\frac{i}{2\left(n +\tfrac{1}{2}\right)} \left(\left( n + \tfrac{1}{2} \right)^{2} -m^{2}\right) ,
\end{align}
for its component $\Psi_1$ and
\begin{align} \label{e: QNF Witten Dirac 2}
\omega_{2} = - \frac{i}{2n} \left(n^{2} - m^{2} \right), \qquad \qquad
\omega_{2} = - \frac{i}{2\left(n+\frac{1}{2}\right)} \left(\left(n + \tfrac{1}{2} \right)^{2} -m^{2} \right),
\end{align}
for its component $\Psi_2$.

Notice that for the Dirac field and the coupled to scalar curvature Klein-Gordon field with $\zeta \leq 1/4$ the real parts of their QNF are equal to zero, while for the coupled to scalar curvature Klein-Gordon field with $\zeta > 1/4$ the real parts of its AQNF depend on the coupling constant $\zeta$, the mass and the mode number of the field. Thus we get that  original Hod's conjecture yields several results for the entropy quantum of the Witten black hole and notice that some of these values depend on the coupling constant and parameters of the field as the mass and the mode number.

For the four sets of exact QNF (\ref{e: QNF 2D Witten KG})--(\ref{e: QNF Witten  Dirac 2}) of the Witten black hole, we find that the physical frequencies are equal to $\omega_{p,n} = n/2,$ and therefore $\Delta \omega_p = 1/2$. Also for this black hole the surface gravity is $\kappa = 1/2$, therefore from these expressions we obtain that $\Delta \omega_p = \kappa$. Following the procedure explained in Eqs.\ (\ref{e: adiabatic invariant 2D})--(\ref{e: spectrum discrete}) for the black holes (\ref{e: black holes 2D}), we find that the Hod-Maggiore proposal produces that the entropy quantum of the Witten black hole is equal to $\Delta S = 2 \pi$. We notice that this value for $\Delta S$ is independent of the coupling constant and the properties of the field.

In contrast to the original Hod conjecture, for the Witten black hole the Hod-Maggiore proposal yields one value for $\Delta S$ and it is independent of field's properties, moreover, it is equal to the entropy quantum of the black holes (\ref{e: black holes 2D}).

Now we focus on the AdS$_2$ black hole \cite{Achucarro:1993fd}. As we commented, the Hod-Maggiore proposal yields a puzzling result for the entropy quantum of the $D$-dimensional Schwarzschild anti-de Sitter black hole \cite{Daghigh:2008jz,Rong:2010,Wei:2009sw}. Since the AdS$_2$ black hole is asymptotically anti-de Sitter, we believe that it is convenient to compute its entropy spectrum. In what follows we take the metric of the AdS$_2$ black hole as  \cite{Achucarro:1993fd} 
\begin{equation} \label{e: metric AdS(2) black hole}
\dd s^{2} = - \left(-M + \frac{x^{2}}{L^{2}}\right) \dd t^{2} + \left(-M + \frac{x^{2}}{L^{2}}\right)^{-1} \dd x^{2},
\end{equation}
where $M$ is the mass of the black hole, $L$ is related to the cosmological constant, and the radius of the horizon is given by $x_H = \sqrt{M} L$. We notice that for this black hole its surface gravity $\kappa$ is equal to
\begin{equation} \label{e: surface gravity AdS(2)}
 \kappa =  \frac{\sqrt{M}}{L}.
\end{equation} 

For the AdS$_2$ black hole the QNF of the massive Klein-Gordon and Dirac fields are calculated exactly in Ref.\ \cite{ALO-RCE-IVA}. For the  Klein-Gordon field its QNF are
\begin{align} \label{e: QNF AdS(2) Klein Gordon}
 \omega = -i \frac{\sqrt{M}}{L} \left(2 n + \tfrac{1}{2} + \tfrac{1}{2}\sqrt{1 + 4 m^{2}} \right), \quad
 \omega = -i \frac{\sqrt{M}}{L} \left(2 n + \tfrac{3}{2} + \tfrac{1}{2}\sqrt{1 + 4 m^{2}} \right),
\end{align}
with $n=0,1,2,\dots$. While for the Dirac field the QNF  are
\begin{equation} \label{e: QNF Dirac 1 AdS(2)}
 \omega_1 = -i \frac{\sqrt{M}}{L} \left(2n + m + \tfrac{3}{2} \right),  \qquad \qquad \omega_1 = -i \frac{\sqrt{M}}{L} \left(2n+ m + \tfrac{5}{2} \right),
\end{equation} 
for its component $\Psi_1$ and
\begin{equation} \label{e: QNF Dirac 2 AdS(2)}
 \omega_2 = -i \frac{\sqrt{M}}{L} \left(2n + m + \tfrac{1}{2} \right),  \qquad \qquad \omega_2 = -i \frac{\sqrt{M}}{L} \left(2n + m + \tfrac{3}{2} \right),
\end{equation} 
for its component $\Psi_2$.

From these expressions for the QNF of the Klein-Gordon and Dirac fields, we find that the physical frequency of the AdS$_2$ black hole is
\begin{equation}
 \omega_{p,n} = 2 \frac{\sqrt{M}}{L} n = 2 \kappa n .
\end{equation} 
Hence we obtain that $\Delta \omega_p = 2 \kappa $ for the AdS$_2$ black hole. 

Using the second method we get that for the AdS$_2$ black hole, the Hod-Maggiore proposal produces an entropy quantum equal to 
\begin{equation} \label{e: quantum ads2}
\Delta S = 4 \pi .
\end{equation}
We notice that this entropy quantum is equal to that of the static BTZ black hole \cite{Kwon:2010um} and the $D$-dimensional MTBH \cite{LopezOrtega:2010wx} (and of the $D$-dimensional de Sitter spacetime \cite{LopezOrtega:2009ww}). 

We believe that the reason for this result is that the AdS$_2$ black hole, the static BTZ black hole, and the $D$-dimensional MTBH have metrics with identical $(t,x)$ sector  given by (up to a conformal transformation with a constant function and a time rescaling)
\begin{equation} \label{e: metric sector tr}
 \dd s^2 =  - (x^2 -x_+^2) \dd t^2 + (x^2 - x_+^2)^{-1} \dd x^2 ,
\end{equation} 
with $x_+$ denoting a constant that determines the event horizon. This fact points out that for these black holes only the sectors $(t,x)$ of their metrics are relevant in the determination of  their entropy quanta with the Hod-Maggiore proposal. We believe that this point deserves further research.

\section{Comments on previous results}
\label{s: Some extensions}

For the AQNF of single horizon black holes it is common to find the behavior \cite{Nollert:1993zz}--\cite{Natario:2004jd}, \cite{LopezOrtega:2006my,LopezOrtega:2011sc,ALO-RCE-IVA}: (i) Their real parts go to a constant; (ii) Their imaginary parts behave as $|\omega_{I,n}| \approx n \kappa q$. For example, for the AQNF of the gravitational perturbations moving in the four-dimensional Schwarzschild black hole \cite{Hod:1998vk}, \cite{Nollert:1993zz}--\cite{Natario:2004jd} we obtain that $\omega_{R,n} = \ln (3)/ 8 \pi M$ and $\omega_{I,n} = n \kappa q$ with  $q=1$ (see the formula (\ref{e: Schwarzschild AQNF 4D})). Making these assumptions we get that the imaginary parts of the AQNF give the main contribution to the physical frequencies (\ref{e: definition physical frequency}), and therefore
\begin{equation} \label{e: Delta omega kappa}
  \Delta \omega_p = \kappa q .
\end{equation}

Considering the Hod-Maggiore proposal, from the formula (\ref{e: delta E}), and the first law of the thermodynamics we find that
\begin{equation}
 \Delta S = \frac{\hbar \kappa }{T} q .
\end{equation}  
Taking $T$ as the Hawking temperature (\ref{e: Hawking temperature}), we get that for single horizon black holes such that their AQNF satisfy the previous conditions, the Hod-Maggiore proposal gives that their entropy quanta are equal to
\begin{equation} \label{e: entropy quantum linear}
\Delta S = q(2 \pi).
\end{equation}   
Assuming the previous conditions on $\omega_{R,n}$ and $\omega_{I,n}$ we also get this value for $\Delta S$ with the modified Kunstatter method. We notice that based on dimensional arguments, the first law, and Hod's conjecture, Kunstatter \cite{Kunstatter:2002pj} obtains an equally spaced spectrum for the Schwarzschild black hole. In Ref.\ \cite{Kunstatter:2002pj} it is identified the denominator of Eq.\ (\ref{e: adiabatic invariant}) with the real part of the AQNF. Here, following to Maggiore, to get the entropy quanta (\ref{e: entropy quantum linear}) we take the denominator of Eq.\ (\ref{e: adiabatic invariant}) equal to the quantity $\Delta \omega_p$ \cite{Vagenas:2008yi,Medved:2008iq}.

From this result it is straightforward to determine the value $\Delta S = 4 \pi$ obtained for the entropy quantum of the AdS$_2$ black hole, the three-dimensional static BTZ black hole \cite{Kwon:2010um}, and the $D$-dimensional MTBH  \cite{LopezOrtega:2010wx}, because for these three black holes we get $\Delta \omega_p = 2 \kappa $ (for the QNF of the Klein-Gordon and Dirac fields given above, for the QNF of the Klein-Gordon field used in Ref.\ \cite{Kwon:2010um}, and for the QNF of the gravitational perturbations exploited in Ref.\ \cite{LopezOrtega:2010wx}). Taking into account the formula (\ref{e: entropy quantum linear}) we obtain $\Delta S = 2 (2 \pi) = 4 \pi$, as we computed in the formula (\ref{e: quantum ads2}) and as calculated in Refs.\ \cite{Kwon:2010um} and \cite{LopezOrtega:2010wx} (see below).

Thus, in the framework of the Hod-Maggiore proposal, to determine the entropy quantum $\Delta S$ of a single horizon black hole, we must calculate the quantity $\Delta \omega_p$ as a function of the surface gravity $\kappa$, and if the relation between them is linear, as in the expression (\ref{e: Delta omega kappa}), then the entropy quantum takes the form (\ref{e: entropy quantum linear}). As far as we know this assertion is not published.

It is known that there are spacetimes such that the QNF (or at least the AQNF) of several fields are purely imaginary, that is, as $n \to \infty$, we find that $\omega_{R,n} = 0$ \cite{Natario:2004jd,LopezOrtega:2006my,LopezOrtega:2005ep}. For these spacetimes it is appropriate to determine the entropy quantum that gives original Hod's conjecture \cite{Hod:1998vk,Maggiore:2007nq}.

According to Hod's conjecture, the formula (\ref{e: Hod conjecture formula}) determines the change of energy for the black hole when it emits a quantum, hence in the limit $\omega_{R,n} \to 0$ we obtain that $\Delta E = 0$. If the first law of the thermodynamics is valid for the event horizons of these black holes that is, the change in the horizon entropy and the variation in the energy satisfy
\begin{equation} \label{e: first law}
 \Delta E = T \Delta S,
\end{equation}
where $T$ is the temperature associated with the horizon, we find that for $T \neq 0$\footnote{We obtain a similar result with the Kunstatter method of Ref.\ \cite{Kunstatter:2002pj}.} 
\begin{equation}
 \lim_{\omega_{R,n} \to 0} \Delta S = 0, 
\end{equation} 
and hence for the spacetimes whose QNF (or at least their AQNF) are purely imaginary, the Hod conjecture predicts that the emission of a quantum does not produce a change in their horizon entropies.\footnote{We note that this conclusion is also valid when the changes in the horizon entropy and the energy of the black hole are related by 
\begin{equation}
 \Delta S = \gamma_1 (\Delta E) + \gamma_2  (\Delta E)^2 + \gamma_3  (\Delta E)^3 + \dots
\end{equation} 
where the $\gamma_i$ are constants different from zero.} Hod \cite{Hod:1998vk} and Maggiore \cite{Maggiore:2007nq} comment on this fact, but it is not proved in detail. We state in explicit form the previous fact because it will be useful for the discussion that follows.

The result applies to the AdS$_2$ black hole, because its QNF are purely imaginary. This similarly applies to the Witten black hole, because some of its QNF are purely imaginary. Therefore for these fields the Hod conjecture produces that the entropy of the horizon does not change when the black hole radiates a quantum.

Up to this point, we calculate the entropy spectra of the black holes (\ref{e: black holes 2D}) supported on the AQNF of the Klein-Gordon field, and for the Witten and AdS$_2$ black holes we exploit the exact QNF of the Klein-Gordon and Dirac fields. Based on the different status of the gravitational perturbations and other perturbing fields in the $D$-dimensional black holes with $D \geq 4$, in Ref.\ \cite{LopezOrtega:2010wx} (see also Ref.\ \cite{Andersson:2003fh}) it is suggested that in the Hod-Maggiore proposal we must use the AQNF of the gravitational perturbations to calculate the value of the entropy quantum, because these are perturbations in the structure of the spacetime itself and therefore it is more probable that they carry information about the structure of the horizon. This suggestion is in contrast to those by Hod \cite{Hod:1998vk} and Maggiore \cite{Maggiore:2007nq}, who assume that we can use the AQNF of other classical fields to calculate the entropy quantum.

To determine the more appropriate option, we take into account the physical picture of the process that produces the change of the horizon entropy  in the Hod \cite{Hod:1998vk} and the Hod-Maggiore \cite{Maggiore:2007nq} proposals. According to Hod, the emission of a quantum  yields a variation in the energy of the black hole given by the formula (\ref{e: Hod conjecture formula}) (or by the expression (\ref{e: delta E}) according to Maggiore). In this process we do not restrict the emission of quanta only to the gravitational perturbations, because the black hole can radiate other fields, for example, the Klein-Gordon, Dirac, and electromagnetic fields \cite{Fabbri Navarro book}.

Hence, in the frameworks of the Hod and the Hod-Maggiore proposals, it is reasonable to assume that for a field different from the gravitational perturbation, the emission of its quantum yields a change in the horizon entropy determined by its AQNF and not by the AQNF of the gravitational perturbations. 

Thus, in the Hod and Hod-Maggiore proposals, we can determine the variation in the horizon entropy due to emission of a quantum based on the AQNF of fields different from the gravitational perturbations, and a more appropriate question is if the change in the horizon entropy depends on the emitted field. Hence we calculate the change in the horizon entropy that produces the emission of a quantum of a field different from the gravitational perturbation.  

For the backgrounds analyzed in Refs.\ \cite{Maggiore:2007nq,LopezOrtega:2009ww,Wei:2009yj,Kwon:2010um,LopezOrtega:2010wx,LopezOrtega:2010tg}, and considering the Hod and Hod-Maggiore proposals, we calculate the change in the horizon entropy that produces the emission of a field different from the gravitational perturbation. We determine whether the change of the entropy depends on the emitted field and its parameters or the variation in the horizon entropy is the same for all the studied fields.


\begin{table}[ht]
\caption{AQNF of the $D$-dimensional Schwarzschild black hole and exact QNF of the $D$-dimensional MTBH, $D$-dimensional de Sitter, and  three-dimensional static BTZ spacetimes }
{\begin{tabular}{@{}lcc@{}} \hline
Perturbation &  QNF & $\Delta \omega_p $  \\ 
\hline
&  $D$-dimensional Schwarzschild black hole \cite{Hod:1998vk,Maggiore:2007nq,Wei:2009yj,LopezOrtega:2010tg} &  \\ 
& $\dd s^2 = -\left( 1 - \frac{2 \mu }{r^{D-3}} \right) \dd t^2 +  \left( 1 - \frac{2 \mu }{r^{D-3}} \right)^{-1} \dd r^2 + r^2 \dd \Omega_{D-2}^2$ &  \\ 
\hline
Klein-Gordon \cite{Motl:2002hd,Motl:2003cd,Andersson:2003fh,Natario:2004jd} & $\pm \tfrac{\kappa}{2 \pi} \ln 3 - i \kappa \left( n + \tfrac{1}{2}\right)$ & \\
Gravitational \cite{Motl:2002hd,Motl:2003cd,Andersson:2003fh,Natario:2004jd,Birmingham:2003rf} & $ \pm \tfrac{\kappa}{2 \pi} \ln 3 - i \kappa \left( n + \tfrac{1}{2}\right) $ & $\kappa$ \\
Dirac \cite{Cho:2005yc,Cho:2007zi} & $- i \kappa n$ & \\ 
Electromagnetic \cite{LopezOrtega:2006vn} & $\pm \tfrac{\kappa}{2 \pi} \ln (1 + 2 \cos(\pi j)) - i \kappa \left( n + \tfrac{1}{2}\right) $ &  \\
\hline
 & $D$-dimensional MTBH \cite{LopezOrtega:2010wx} & \\
& $\dd s^2 = - \left(-1 + \frac{r^2}{L^2} \right) \dd t^2 + \left(-1 + \frac{r^2}{L^2} \right)^{-1} \dd r^2  +  r^2 \dd \Sigma_{D-2}^2 $ & \\ \hline
Klein-Gordon \cite{Aros:2002te,Oliva:2010xn} & $ \pm \frac{\xi}{L} - \frac{i}{L}\left(2n + 1  + \sqrt{\left(\tfrac{D-1}{2}\right)^2 + \left(m^2 - \zeta \frac{D(D-2)}{4 L^2}\right)^2 L^2} \right)$ & \\
Gravitational \cite{Birmingham:2006zx} & $\pm  \frac{\xi}{L} - \frac{2i}{L}\left( n + \frac{\mathbb{A}_G}{4} \right)$ & $2 \kappa$ \\
Dirac \cite{LopezOrtega:2010uu} & $-\frac{K}{L} - \frac{i }{L} \left(2 n  +\frac{3}{2} + m L \right) $,\,\,\,\,\, $\frac{K}{L} - \frac{i}{L} \left(2 n + \frac{1}{2} + m L \right) $    &  \\ 
Electromagnetic \cite{Oliva:2010xn,LopezOrtega:2007vu} & $ \pm \frac{\xi}{L} - \frac{2i}{L}\left( n + \frac{\mathbb{A}_E}{4} \right)$ & \\
\hline
&$D$-dimensional de Sitter spacetime \cite{LopezOrtega:2009ww} & \\
&${\rm d}s^2 = - \left(1 -\frac{r^2}{L^2} \right) {\rm d}t^2 +  \left(1 -\frac{r^2}{L^2} \right)^{-1} {\rm d} r^2 + r^2 {\rm d} \Omega^2_{D-2}$ & \\ 
\hline
Klein-Gordon \cite{Natario:2004jd,LopezOrtega:2006my} & $ \pm \frac{i}{L} \left( \frac{(D-1)^2}{4} - (mL)^2 \right)^{1/2}  - \frac{i}{L} \left( l + 2 n +\frac{D-1}{2} \right) $ & \\
& $ \pm \frac{1}{L}  \left( (mL)^2 - \frac{(D-1)^2}{4} \right)^{1/2}  - \frac{i}{L} \left( l + 2 n +\frac{D-1}{2} \right) $ &  \\
Gravitational \cite{Natario:2004jd,LopezOrtega:2006my} & $- \frac{i}{L} (l + D-1-Q + 2 n)$,\,\,\,\,\,  $- \frac{i}{L} (l + Q + 2 n) $ & $2 \kappa$ \\
Dirac \cite{LopezOrtega:2007sr} & $- m - \frac{i}{L}  \left(J + \frac{D+1}{2} + 2 n \right) $,\,\,\,\,\,   $m  - \frac{i}{L} \left(J + \frac{D-1}{2} + 2 n \right)$ & \\
Electromagnetic \cite{Natario:2004jd,LopezOrtega:2006my} & $- \frac{i}{L} (l + D-1-Q + 2 n)$,\,\,\,\,\, $ - \frac{i}{L} (l + Q + 2 n) $ & \\
\hline
& Three-dimensional static BTZ black hole \cite{Wei:2009yj,Kwon:2010um,Setare:2003hm} & \\
& $ \dd s^2 =  - (r^2 -r_+^2) \dd t^2 + (r^2 -r_+^2)^{-1} \dd r^2 + r^2 \dd \phi^2 $ & \\ \hline
Klein-Gordon \cite{Birmingham:2001pj,Cardoso:2001hn,Birmingham:2001hc} & $ \pm k - 2i r_+ \left( n + \tfrac{1}{2}\left(1+ \sqrt{1+m^2}\right) \right) $ & \\
Dirac \cite{Birmingham:2001pj}  &  $ -k - i r_+ \left(2n + \tfrac{3}{2} + m\right) $,\,\,\,\,\,  $k - i r_+ \left(2n + \tfrac{1}{2} + m\right)$  & $ 2 \kappa $  \\
Massive vector field \cite{Birmingham:2001pj}   & $- k - 2 i r_+ \left(n + 1 + \tfrac{m}{2} \right)$,\,\,\,\,\, $k - 2 i r_+ \left(n + \tfrac{m}{2}\right)  $  & \\
\hline
\end{tabular} \label{t: QNF spacetimes}}
\end{table}


To discuss this point, the AQNF of the $D$-dimensional Schwarzschild black hole and the exact QNF of the $D$-dimensional MTBH, the $D$-dimensional de Sitter spacetime, and the three-dimensional static BTZ black hole are listed in Table \ref{t: QNF spacetimes} . 

In Table \ref{t: QNF spacetimes}, as previously, we denote the surface gravity by $\kappa$, the mode number by $n$, and the mass of the field by $m$. For the Schwarzschild black hole $\mu$ is related to the mass of the black hole, $\dd \Omega_{D-2}^2$ is the line element of a $(D-2)$-dimensional sphere. In the AQNF of the electromagnetic field, the parameter $j$ is equal to $j=2(D-3)/(D-2)$  ($j= 2/(D-2)$) for the scalar (vector) type field \cite{LopezOrtega:2006vn}. In the MTBH,  $\dd \Sigma^2_{D-2}$ stands for the line element of a $(D-2)$-dimensional compact space of negative curvature, $\Lambda$ is the cosmological constant, $L^2 = - (D-1)(D-2)/(2 \Lambda)$, $\zeta$ represents the coupling constant between the Klein-Gordon field and the scalar curvature, $\xi$ ($K$) denotes the eigenvalues of the Laplacian (Dirac) operator on the $(D-2)$-dimensional base manifold, $\mathbb{A}_G=\mathbb{A}_E = D-1$ ($\mathbb{A}_G=\mathbb{A}_E = |D-5|+2 $) for the gravitational and electromagnetic perturbations of vector (scalar) type, and $\mathbb{A}_G= D+1$ for the tensor type gravitational perturbations ($D\geq5$). For the de Sitter spacetime, $L$ is the curvature radius and it is related to the cosmological constant $\Lambda$ by the expression $ L^2 = (D-1)(D-2)/2 \Lambda$, $l$ ($J$) are the eigenvalues of the Laplacian (Dirac) operator on the $(D-2)$-dimensional sphere, $Q = 1$ ($Q=2$)  for the gravitational and electromagnetic perturbations of vector (scalar) type, and $Q=0$ for the gravitational perturbations of tensor type. Finally, for the static BTZ black hole, $r_+$ denotes the radius of the event horizon and $k$ is the azimuthal number.

From the QNF listed in Table \ref{t: QNF spacetimes} we see that for the Schwarzschild and de Sitter spacetimes the real parts of their AQNF may be equal to zero or different from zero. For the four spacetimes of Table \ref{t: QNF spacetimes} when the real parts are different from zero, they depend on characteristics of the field as the mass and the mode. Also for the Schwarzschild and de Sitter spacetimes the real parts of their AQNF may depend on the dimension. 

From our previous results we state that for the fields whose AQNF are purely imaginary, the Hod conjecture produces that the horizon entropy does not change when a quantum of this field is radiated. Furthermore for the four backgrounds of Table \ref{t: QNF spacetimes}, when the real parts of the AQNF are different from zero the Hod conjecture yields that the horizon entropy change in discrete steps and their values depend on the properties  as the mass and the mode number of the field exploited in the calculation (also these may depend the eigenvalues of the Laplace and Dirac operators on the base manifold in MTBH and BTZ black holes and on the spacetime dimension in Schwarzschild and de Sitter backgrounds). For example, according to Hod's conjecture, in de Sitter spacetime the emission of a quantum of the massive Klein-Gordon field changes the horizon entropy for $m^2 > (D-1)^2/L^2$, whereas for $m^2 < (D-1)^2/L^2$ the horizon entropy remains constant. We find a similar behavior for the Dirac field.

For each spacetime of Table \ref{t: QNF spacetimes}  and for the enumerated fields the imaginary parts of their AQNF behave in a similar way. Notice that in the asymptotic limit they give the main contribution to $\Delta \omega_p$ and as shown in Table \ref{t: QNF spacetimes} we get only a result for this quantity. For the spacetimes of Table \ref{t: QNF spacetimes}, the Hod-Maggiore proposal produces that when they emit a quantum, the change in the horizon entropy does not depend on the radiated field (or the spacetime dimension). From the previous results and using the data of Table \ref{t: QNF spacetimes} we get that for the Schwarzschild black hole the entropy quantum is equal to $\Delta S = 2 \pi$, while for the MTBH, the de Sitter spacetime, and the BTZ black hole the entropy quantum is equal to $\Delta S = 4 \pi$. Thus for these spacetimes the Hod-Maggiore proposal gives more general results than the original Hod conjecture.

For the spacetimes enumerated in Table \ref{t: QNF spacetimes}, our results for their entropy quanta extend those of Refs.\   \cite{Maggiore:2007nq,LopezOrtega:2009ww,Wei:2009yj,Kwon:2010um,LopezOrtega:2010wx,LopezOrtega:2010tg}, where to get the results for the entropy quanta of the Schwarzschild black hole, the MTBH, and the de Sitter spacetime only the QNF of the gravitational perturbations are exploited and for the BTZ black hole only the QNF of the Klein-Gordon field are used. For the spacetimes with $D \geq 4$ and the fields enumerated in Table \ref{t: QNF spacetimes}, the Hod-Maggiore proposal yields that the emission of a quantum produces the same change in the horizon entropy that the emission of a quantum of the gravitational perturbations produces. For other spacetimes we must verify whether this prediction of the  Hod-Maggiore proposal is valid.

Thus we can determine the entropy spectra of the two-di\-men\-sion\-al black holes (\ref{e: black holes 2D}), (\ref{e: metric Witten}), and (\ref{e: metric AdS(2) black hole}) based on the AQNF of the Klein-Gordon field or Dirac field, because we calculate the variation of the horizon entropy that produces the emission of a quantum of the Klein-Gordon field or the Dirac field (also we must recall that in two dimensions (and in three dimensions) the gravitational field does not possess propagating degrees of freedom, and therefore in two-dimensional (and in three-dimensional) spacetimes we cannot restrict to  the AQNF of the gravitational perturbations). Notice that for the Witten and AdS$_2$ black holes the Hod-Maggiore proposal yields that the changes in their horizon entropies do not depend on the emitted field (for Klein-Gordon or Dirac fields).

\section{Summary}
\label{s: Summary}

In short, for the entropy quanta of the Witten black hole, the AdS$_2$ black hole, and the black holes (\ref{e: black holes 2D}), we show that the Hod-Maggiore proposal produces more general results than the original Hod conjecture. For the entropy quanta of the Witten black hole and the black holes (\ref{e: black holes 2D}) we obtain one value, $\Delta S = 2 \pi$ and it is equal to the value found for several higher dimensional asymptotically flat black holes as Schwarzschild, Kerr, and Reissner-Nordstr\"om. Moreover for the AdS$_2$ black hole we get that the Hod-Maggiore proposal produces an entropy quantum equal to $\Delta S = 4 \pi$, which is identical to the value calculated for other asymptotically anti-de Sitter black holes as the three-dimensional BTZ black hole and the $D$-dimensional MTBH. Notice that for the AdS$_2$ black hole the original Hod conjecture gives that the emission of a quantum does not produce a change in its horizon entropy.

In this paper, for the spacetimes studied and for the fields analyzed, we show that the Hod-Maggiore proposal predicts that the emission of a quantum produces the same change in the entropy of the horizon, in contrast, the original Hod conjecture yields that the variation of the horizon entropy depends on the properties of the emitted field and of the spacetime. Hence for the spacetimes studied in the previous sections we find that the values predicted by the Hod-Maggiore proposal for their entropy quanta  are more general than the results given by the Hod conjecture.

Moreover, our results point out that in the Hod and Hod-Maggiore proposals we can use the AQNF of different fields and hence we must not restrict to the AQNF of the gravitational perturbations as suggested in Refs.\ \cite{Andersson:2003fh}, \cite{LopezOrtega:2010wx}. So we believe that in the framework of the Hod and Hod-Maggiore proposals it is more appropriate to ask whether the change in the horizon entropy is independent of the radiated field. For the black holes that we study in the previous sections we showed that the emission of a quantum yields the same change in the horizon entropy (at least for the fields of Table \ref{t: QNF spacetimes}) and we believe that this fact is a significant prediction of the Hod-Maggiore proposal. Thus for those black holes in $D \geq 4$ whose entropy quanta are calculated based on the AQNF of a field different from the gravitational perturbation  \cite{Setare:2010gi,Wei:2010yx,Gonzalez:2010vv,Chen:2010cp}, we must use the AQNF of the gravitational perturbations to compute the entropy quanta and compare the results.

From the previous results and for the black holes  analyzed here, we see that the Hod-Maggiore proposal produces that the entropy quantum is different for asymptotically flat and asymptotically anti-de Sitter black holes. Thus in general relativity with cosmological constant this proposal gives at least three different values for the entropy quantum of single horizon black holes ($\Delta S  = 2\pi$, $\Delta S  = 4 \pi$, and $\Delta S = 4 \pi \sin( \pi/(D-1))$) and not a universal value. As far as we know, in general relativity, for asymptotically flat, single horizon black holes the Hod-Maggiore proposal yields $\Delta S  = 2\pi$. Notice that if the proposal of Medved \cite{Medved:2009nj,Medved:2010iy} is true, and there is a universal entropy quantum, then we must discard the Hod and Hod-Maggiore proposals.

Recently there is a discussion on what spectrum (entropy or area) of a horizon is more fundamental \cite{Wei:2009yj,Kothawala:2008in,Ren:2010,Kwon:2010km}. For the studied black holes in this work we do not consider similar issues, but notice that we only calculate their entropy spectra. For the Schwarzschild, MTBH, BTZ, and de Sitter spacetimes it is straightforward to compute their area spectra because for them the Bekenstein-Hawking formula is valid. Furthermore taking into account the results of Ref.\ \cite{Grumiller:2007ju} we see that for the two-dimensional Witten, AdS$_2$, and single horizon black holes (\ref{e: black holes 2D}), the following relation is valid 
\begin{equation} \label{e: Bekenstein Hawking}
 S = \frac{A}{4 G_{e}} ,
\end{equation} 
where $A$ is the area of the event horizon, and $G_{e}=G / \phi_h$, is the effective Newton constant. So for these two-dimensional black holes, it is valid that the entropy is equal to one quarter  of the area in ``effective Planck units'' \cite{Grumiller:2007ju}. Therefore from formula (\ref{e: Bekenstein Hawking}) and the results for the entropy spectra we can calculate the area spectra.

\section{Acknowledgments}

This work is supported by CONACYT M\'exico, SNI M\'exico, EDI-IPN, COFAA-IPN, and Research Projects SIP-20110729 and SIP-20111070.

\end{document}